\documentclass[aps,pre,preprint,groupedaddress]{revtex4}

\usepackage{graphicx}
\usepackage{dcolumn}
\usepackage{bm}
\usepackage{array}
\usepackage[large]{subfigure}

\begin{document}

\title{Self-learning Multiscale Simulation for Achieving High Accuracy and High Efficiency Simultaneously}
\author{Wenfei Li and Shoji Takada}
\email{takada@biophys.kyoto-u.ac.jp}
\affiliation{Department of Biophysics, Graduate School of Science, Kyoto University,
Kyoto 606-8502, Japan, CREST, Japan Science and Technology Inc, Japan.}
\date{\today}

\begin{abstract}
We propose a new multi-scale molecular dynamics simulation method which can
achieve high accuracy and high sampling efficiency simultaneously without
aforehand knowledge of the coarse grained (CG) potential and test it for a
biomolecular system. Based on the resolution exchange simulations between
atomistic and CG replicas, a self-learning strategy is introduced to
progressively improve the CG potential by an iterative way. Two tests show
that, the new method can rapidly improve the CG potential and achieve
efficient sampling even starting from an unrealistic CG potential. The resulting
free energy agreed well with exact result and the convergence by the method
was much faster than that by the replica exchange method. The method is generic
and can be applied to many biological as well as non-biological problems.

\end{abstract}

\pacs{87.15.A-, 87.10.Tf}
\maketitle



\section{INTRODUCTION}
Biomolecular systems, and more broadly soft matters, are inherently
hierarchic: Atomic details are crucial for functioning, which is often
regulated by slower and larger-scale motion. Not surprisingly, both all-atom
(AA) detailed and coarse-grained (CG) molecular dynamics (MD) simulation,
and sometimes Monte-Carlo simulation,
methods have been developed and they play more and more crucial roles in
biophysics\cite{LE75,KA02,TO05,MC77,TA99,LI08,GI06}, as complement of experiments. Unfortunately, due to the large
number of degrees of freedom involved and inherently rugged energy surface,
the time scale currently reachable by the AA simulation, $\sim$microsecond,
is far below typical biologically relevant time scale of milliseconds or
longer. Conversely, the CG simulation can sample molecular conformations
much more efficiently, but it is unavoidably less accurate in energy
estimation. Thus, the AA (CG) simulations are more (less) accurate in
energy, but less (more) efficient in sampling. To surmount these problems, a
number of strategies have been proposed to integrate the AA and CG
simulations, which is often called multiscale simulations\cite{AY07}.

In one widely used multiscale strategy, first, the conformational space is
broadly sampled by a certain CG simulation. The sampled CG ensemble is then
converted to AA detailed ensemble followed by some refinement. This CG$%
\rightarrow $AA strategy has been successfully used in protein folding and
structure prediction\cite{HE07,KM08,BR05}. Recently, a more parallel method with
the two-way coupling, called the resolution replica exchange (ResEx), was
proposed in the framework of Hamiltonian replica exchange MD\cite{SU00,FU02,KW05}, in which the AA MD is
coupled to a certain CG MD via trials of conformation exchange\cite%
{LY06a,LY06b,LI07,CH06,LW05}. Both of these multiscale methods, however, rely heavily on
the accuracy of the CG model. If the CG potential has its major basins
different from those of AA potential, neither the CG$\rightarrow $AA
strategy nor the ResEx works well. In practice, however, as noted above, the
CG model is unavoidably less accurate. Therefore, the inaccuracy of the CG
potential represents the major bottleneck of the multiscale simulations\cite%
{CH06a}. Multiscale simulation methods which do not depend on the aforehand
knowledge of the accurate CG potential, is desired and one such method is
developed in this work.

Often the CG potential can be derived by yet another multiscale approach of
AA$\rightarrow $CG type. Performing the AA simulation, we can use
energy/force of the AA model to estimate the effective energy/force acting
on CG particles\cite{CH06b,IZ05,ZH07,BA07,RE03,TR05,MO07,MA07}. However, as the AA
simulation can sample very limited conformational space, the extracted CG
potential well-approximates the AA potential in narrow conformational space.
In particular, as in many practical applications such as protein structure
prediction or protein-protein docking, we do not know relevant part of the
conformational space a priori, and so the AA simulation cannot reach the
relevant part of the phase space and thus the extracted CG potential is
useless in such cases.

In this work, we report a new multiscale simulation method, self-learning
multiscale molecular dynamics (SLMS-MD), in which the CG potential is
continuously improved according to the previously sampled CG conformations
and their corresponding AA energies by an iterative way. The CG simulation
ensures the efficient and broad sampling, and simultaneously the AA energies
shape up the accuracy of the CG potential. The most promising feature of
this method is that its performance does not rely on the accuracy of the
initial CG potential because the CG potential is progressively improved by
self-learning. Two kinds of test studies demonstrate that by using the
SLMS-MD, we can optimally combine the advantages of the AA and CG
simulations, and achieve high accuracy and high sampling-efficiency
simultaneously. To our knowledge, this is the first work that tightly
couples the CG sampling and AA energy to improve the CG potential, and
realize efficient and accurate multiscale simulations.

\section{THEORIES AND METHODS}
Fig. 1 shows the detailed flow chart of the SLMS-MD. In this method, we start with an arbitrarily chosen CG
potential and the AA potential that is assumed to be accurate. Given two potentials, we
perform ResEx MD simulations\cite{LY06a}. Namely, two independent simulations with CG and AA resolutions are
conducted simultaneously in parallel at temperatures of $T_{CG}$ and $T_{AA}$, respectively. After certain MD
steps, exchange of conformations of these two replicas is attempted. Here we assume that we have a method of
structure mapping between AA and CG representations, detail of which depends on cases and will be described
later. The acceptance ratio of the exchange between
the CG replica with coordinate $X_{i}$ and the AA replica with coordinate $%
X_{j}$ is determined by Metropolis-like criterion:
\begin{equation}
P(X_{i}{\leftrightarrow}X_{j})=\min (1,\exp (-\Delta _{AC}))
\end{equation}
where
\begin{equation}
\Delta _{AC}=\beta
_{CG}(E_{CG}(X_{j})-E_{CG}(X_{i}))+\beta _{AA}(E_{AA}(X_{i})-E_{AA}(X_{j}))
\end{equation}
with $\beta _{CG}=1/(k_{B}T_{CG})$ and $\beta _{AA}=1/(k_{B}T_{AA})$. $E_{CG}$
and $E_{AA}$ are the corresponding CG energy and AA energy, respectively. After
this ResEx MD simulation, we collect structures sampled by the CG replica,
map them to the AA representations, and compute their AA energies. With
these, the pairwise distribution function (PDF) $g(r)$ of each pair of
interacting CG particles at temperature $T_{AA}$ can be calculated by
re-weighting\cite{FE89}:
\begin{equation}
g(r)=\frac{\sum\limits_{i}{\delta (r_{i}-r)}\exp {(\beta _{CG}E_{CG}(X}_{i}){%
-\beta _{AA}E_{AA}(X}_{i}{))}}{\sum\limits_{i}{\exp (\beta _{CG}E_{CG}{(X}%
_{i})-\beta _{AA}E_{AA}{(X}_{i}))}}
\end{equation}
where $r_{i}$ is the distance for certain pair of CG particles of the $i$-th
structure. We note that this PDF reflects, via re-weighting, energetic
information from the AA potential. The corresponding potential of mean force
(PMF) $w(r)$ can be derived by the standard Boltzmann inversion
\begin{equation}
\beta _{CG}w(r)=-\ln (g(r)/g_{R}(r))
\end{equation}
where the $g_{R}(r)$ is the PDF of the reference state and determined according to the DFIRE method\cite{ZH04a}.
This PMF is further iteratively adjusted to reproduce the original $g(r)$ by CG simulation, which results in the
effective interaction $U_{CG}(r)$ for each pair of interacting CG
particles\cite{RE03,TR05}. Note that due to the iteration, the final $%
U_{CG}(r)$ does not depend on the choice of the $g_{R}(r)$. This $U_{CG}(r)$
is then used as the CG potential of the next-generation ResEx simulation.
Since the $U_{CG}(r)$ is extracted by combining the energy information of
the AA simulation and the conformational information of the CG simulation,
namely, the advantages of both methods, it is expected to be improved compared
with the initial CG potential. In this way, we can derive an accurate CG potential,
and therefore overcome the bottleneck of the multiscale simulations. Such a high
precision CG potential not only ensures meaningful energetics and sampling in the
CG simulation, but also makes the conformation exchange between the CG and AA
replicas more probable, which can speed up the AA MD sampling significantly based on
the ResEx method. Therefore, the SLMS-MD can accomplish high efficiency and high accuracy
simultaneously in both the CG and AA levels.

It is worth noting that although the above described multiscale protocol uses the
re-weighting and Boltzmann inversion method in the self-learning stage, however, the
essential idea of this SLMS-MD does not depend on the specific learning method.
Other learning methods, e.g., the force match protocol\cite{IZ05,ZH07}, should also be applicable.

\begin{figure}[h]
\centering
\includegraphics[width=0.45\textwidth,clip]{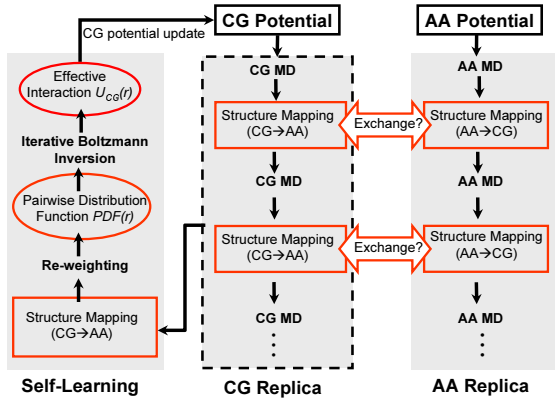}
\caption{Flow chart of the SLMS-MD}
\end{figure}

\section{RESULTS AND DISCUSSIONS}
\subsection{One dimensional system}

We first test the SLMS-MD in an one-dimensional toy model system. The high
resolution(HR) system is represented by a potential function $%
U_{HR}(x)=0.2x^{2}-2.5\cos (4\pi x)$ where the ruggedness modelled by the
cosine mimics that of the AA energy surface (Black line in Fig. 2(a)). As a
control, performing a Langevin MD at the temperature $T=0.5$, we found that
it hardly moves from the initial valley leading to very poor sampling. To
test SLMS-MD, we introduce a CG potential, a simple harmonic potential $%
U_{CG}(x)=k(x-x_{0})^{2}$. Apparently, $k\simeq 0.2$ and $x_{0}=0$, are the
best parameters to approximate the $U_{HR}(x)$. Assuming that we do not a
priori know the right parameters of the CG potential, we started with $k=0.5$
and $x_{0}=6.0$ (Red line in Fig. 2(a)), far from the best values. We
will show that by using the SLMS-MD, we can derive the best CG potential parameters,
with which we can then improve the sampling of the HR system by ResEx method.

From the above initial condition, we conducted the SLMS-MD simulations at $T=0.5$ for both replicas
by Langevin dynamics with a friction coefficient $\gamma =1.0$. The time
steps used were 0.0002 and 0.002, respectively, for the HR and CG replicas.
Every 100 MD steps, exchange of conformations was attempted. After $10^{6}$
exchange attempts, the position probability distribution of the CG
simulation was calculated and mapped to the HR case by re-weighting\cite%
{FE89}. In the first generation (iteration 0), discrepancy between two
energies led to nearly no exchange between replicas as shown in Fig. 2(c).
The resulting position probability distributions of the HR replica
(red line of Fig. 3(a)) and the re-weighted CG replica (red line of Fig. 3(b))
were, as expected, very poor, and deviate far from the theoretical position
probability distribution which is calculated by numerical integration
(black dashed lines in Fig. 3(a) and (b)). Using sampled data, we then
calculated the $w(x)$ and fitted it with a harmonic function. For this
one-dimensional case, the PMF is equivalent
to the effective interaction. The fitted parameters $k$ and $x_{0}$ were
used directly for the next ResEx MD simulation. The above procedure was repeated
until the parameters of the new $U_{CG}$ get converged. Fig. 2(b) shows the
fitted parameters $k$ and $x_{0}$ as a function of learning iteration step.
The resulted CG potential curves after one, two and three steps of iterations
are also plotted in Fig. 2(a). One can see that the CG potential is improved
within very limited steps of learning iterations based on the SLMS-MD. After two steps,
the obtained CG potential fits the  overall behavior of the HR potential very well (Fig. 2(a)).
Due to the best matching between the HR and CG potentials, the maximal acceptance
ratio of ResEx was achieved (Fig. 2(c)) and the sampling quality of the HR
replica was enhanced significantly (Fig. 2(d)). Consequently, after two
steps of iterations, the position probability distribution calculated both
by the HR replica and by the CG replica with re-weighting were almost identical
to the theoretical one (Fig. 3(a) and (b)).

\begin{figure}[h]
\centering
\includegraphics[width=0.45\textwidth,clip]{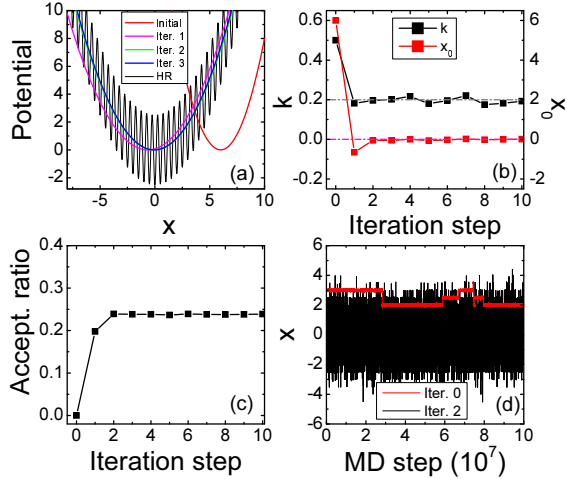}
\caption{(a) CG potentials after one, two and three steps of iterations. The initial CG
potential and the HR potential are also plotted. (b) Parameters of the $U_{CG}$ as
a function of iteration step. (c) Acceptance ratio of the resolution replica exchange
as a function of iteration step. (d) Trajectories of the HR replica before iteration and
after two steps of iterations.}
\end{figure}

\begin{figure}[h]
\centering
\includegraphics[width=0.40\textwidth,clip]{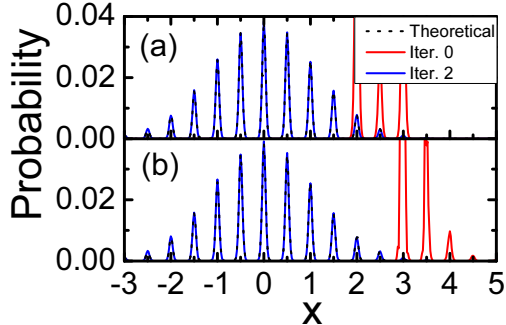}
\caption{Position probability distribution of the HR replica (a) and the
re-weighted CG replica (b) before iteration and after two steps of iterations.
The theoretical distribution which is calculated by
numerical integration is also presented.}
\end{figure}

\subsection{Tri-alanine peptide}

The above one-dimensional toy model clearly demonstrated that the SLMS-MD is highly useful in deriving the
accurate CG potential and accelerating the sampling of the HR system. In this toy model system, the HR potential
can capture the high dimensional characteristics of the AA system through the highly rugged energy function.
However, for real biomolecular systems, the high dimensional characteristics are not only represented by the
highly frustrated energy surface, but also by the entropy contribution, which cannot be captured by the above
one-dimensional toy model system. Therefore, it is important to test the SLMS-MD with a real biomolecule. Here,
we test our method with conformational sampling of a small biomolecule, i.e., a tri-alanine peptide at $T=250K$.
First, a standard temperature replica exchange MD (T-REMD)\cite{SU99,ZH04b} of 60ns with eight replicas from
225.0K to 520.0K gave the well-converged ensemble, with which we can compare our results unambiguously. In
SLMS-MD, the AA replica used an all-atom representation with AMBER force field ff99SB and GB/SA implicit
solvation at $T=250K$ with time step of 0.002ps\cite{AMBER}. The CG replica is represented by $C_{\alpha }$
position. Since the distance between the successive $C_{\alpha }$ does not change significantly, we restrained
it to the equilibrium distance(3.87\AA ). The remaining internal degree of freedom is the distance between the
first and third$C_{\alpha }$ (1-3 $C_{\alpha }$ pair). The initial CG potential between these two residues is
arbitrarily chosen to be $U_{CG}(r)=10.0(r-5.0)^{2}$ where $r$ is the distance between them. After each
iteration, the $U_{CG}(r)$ is updated by the extracted effective interaction which is represented by data table
and interpolated with cubic spline. The CG replica is simulated at temperature of 1.0 using Langevin dynamics
with $\gamma =0.5$ and time step of 0.001.

Both replicas started from extended structures. Every 100 MD steps for the AA replica and $10^{4}$ MD steps for
the CG replica, the conformational exchange was attempted. In mapping the structures from CG replica to AA
replica, software BBQ\cite{GR07} and SCWRL3\cite{CA03} were used. Note that there may be many different possible
AA coordinates for the same CG coordinates. The BBQ and SCWRL just give one solution. The constructed side chain
is the most stable one according to the energy function of SCWRL. Since the energy function of the SCWRL may
deviate from the AA force field we used, it is highly possible that the reconstructed AA structure may not be
accessible by AA simulation at target temperature, therefore the simulation may be irreversible. To overcome
this problem, we correct the produced AA structure by performing equilibrium AA MD simulation at 250K, i.e., the
target temperature, for 1000MD steps. Before this equilibrium MD simulation, the reconstructed AA structure is
heated to 500K to remove possible bad interactions. The equilibrium MD simulation produces a number of possible
AA structures. One of them (the last structure of the equilibrium MD simulation) is used as the starting
structure for AA replica if the exchange is accepted. During the equilibrium MD, the distance between each
$C_{\alpha }$ pair is restrained by a harmonic potential with force constant changing from
10.0kcal/mol/$\mathring{A}^{2}$ to zero gradually. The ResEx MD was performed for 1000 exchange attempts, i.e.,
200ps in AA replica.

After ResEx MD, for the learning stage, all of the sampled CG structures
were converted to AA structures. 2000 steps of equilibrium MD were conducted
with $C_{\alpha }$ distances being restrained by harmonic potential with force
constant of 50kcal/mol/$\mathring{A}^{2}$ after the initial high temperature
simulation. All the sampled structures deposited during this
equilibrium simulation were used to calculate the AA energies for
re-weighting, and then the effective interaction $U_{CG}(r)$ was updated. We
repeated the learning iteration 12 steps.

Fig. 4(a) shows the extracted effective interaction $U_{CG}(r)$ after 3, 6,
9, and 12 steps of iterations, as well as $U_{CG}(r)$ extracted from the
long time T-REMD of 60ns. For comparison, the initial $U_{CG}(r)$ is also
presented in Fig. 4(a)(red line). Clearly, the CG potential was improved rapidly
during the SLMS-MD simulation despite the poor initial potential. After
three steps of iterations, the CG potential already fit the effective
interaction derived from T-REMD very well. The corresponding PDF for the re-weighted
CG replica and the AA replica at different iteration stages were
also calculated and compared with that obtained by T-REMD. Fig. 4 (b) and (c)
shows the PDF of the 1-3 $C_\alpha$ pair calculated by
re-weighted CG replica (b) and AA replica (c), respectively, at different
iteration steps. The target PDF calculated by standard T-REMD is
also plotted in Fig. 4 (b) and (c). Without iteration, the
all-atom simulation cannot give correct PDF. Instead, the distribution was
biased to the position around 5.5\AA\ because of the biased initial CG
potential used, which suggests that if the CG potential is not chosen
appropriately, the exchange between the AA replica and CG replica not only
undermines the sampling efficiency, but may also bias the sampling of the AA
simulation and result in wrong distribution. After three steps of
iterations, both the CG replica and AA replica gave almost identical
distribution to that of the T-REMD, although the ResEx protocol we used may not satisfy
the detailed balance condition strictly\cite{LW05,LY06b}. We emphasize that each step contains
only 200ps MD\ in AA replica.

\begin{figure}[h]
\centering
\includegraphics[width=0.30\textwidth,clip]{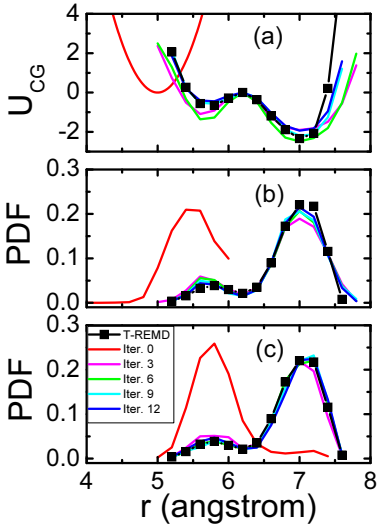}
\caption{CG potential (a) and PDF of the 1-3 $C_\alpha$ pair calculated by
re-weighted CG replica (b) and AA replica (c) at different iteration steps.
The target PDF and CG potential calculated by standard T-REMD are also plotted.}
\end{figure}

A more stringent test of convergence and accuracy is to compare the free energy landscape
on dihedral angles $\phi $ and $\psi $ of the central alanine, as
illustrated in Fig. 5. The result of T-REMD (60ns) shows that it has the
major basin in the extend $\beta $ and poly-proline II conformations, the
second major basin in right-handed (regular) helical conformation, and a
minor basin in the left-handed helices. Between different basins, some
basin-hopping is necessary. After three steps of iterations, the
SLMS-MD(200ps) could reproduce the dihedral angle distribution of the much
longer time T-REMD (60ns) very well. In comparison, the conventional AA MD
(CMD) of 200ps without replica exchange only samples the major basin. The
T-REMD simulation with similar time scale, i.e., 200ps, and the CMD with
much longer simulation time (480ns) sampled both major and second major
basins, but not the minor basin at the left-handed helices. As expected,
without improving the CG potential of the SLMS-MD, both the right-handed
helical and left-handed helical conformations were over populated due to the
poor initial CG potential used.

For a more quantitative test, we compared the ratio of population in the major
basin to that in the second major basin sampled by SLMS-MD and T-REMD. Fig. 6(a)
shows the averaged relative populations between the two major basins calculated
by nine independent SLMS-MD (red line) and T-REMD
(black line) simulations. The error bars are represented by standard deviation.
Although, for both methods, the averaged relative populations come close to
the target value (blue arrow) within $\sim200ps$, the standard deviation (error bar)
for the SLMS-MD is about an order smaller than that of the T-REMD around $%
200ps$, suggesting better sampling convergence in SLMS-MD. Fig. 6(b) also shows the representative
T-REMD (black line) and CMD (red line) trajectories with long time scale (60 ns).
One can see that it needs $\sim 8ns$ for the T-REMD to get well converged. For the CMD
simulation, as expected, the convergence needs much longer time.

The above result shows that the SLMS-MD has higher sampling efficiency compared with the T-REMD even for
this tri-alanine peptide. However, such comparison is based on the AA MD time. For the SLMS-MD,
the self-learning stage, including structure mapping, short equilibrium MD simulation, iterative
Boltzmann inversion, etc., also takes CPU time. Taking these processes into account,
the practical CPU time used by the SLMS-MD for this tri-alanine peptide is even longer than
that used by the T-REMD. This is because the tri-alanine used here is so short that the characteristic
time scale for the converged sampling is shorter than the overhead time needed to implement the self-learning processes.
In the case of larger systems, e.g., proteins with more than 100 amino acids which are the main
interesting targets for most multiscale simulations, the characteristic time
scale for the converged sampling will be overwhelmingly longer than the time needed for implementing
the self-learning processes. In such case, the time for AA MD dominates the entire SLMS-MD process. Thus comparison
by the required time scale of AA MD would be reasonable.

\begin{figure}[h]
\centering
\includegraphics[width=0.45\textwidth,clip]{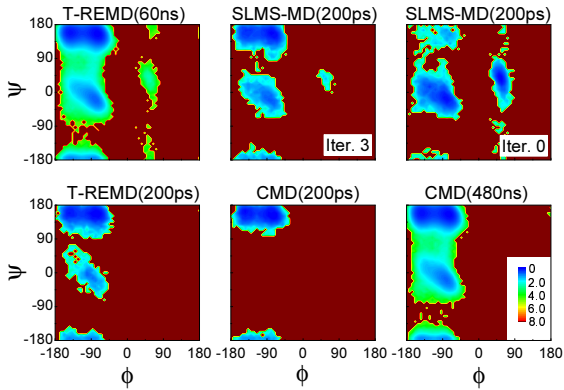}
\caption{Free energy landscape on $\protect\phi $ and $\protect\psi$
dihedral angles of the central alanine at $250K$ calculated with different
protocols. The unit of the free energy is kcal/mol.}
\end{figure}
\begin{figure}[h]
\centering
\includegraphics[width=0.35\textwidth,clip]{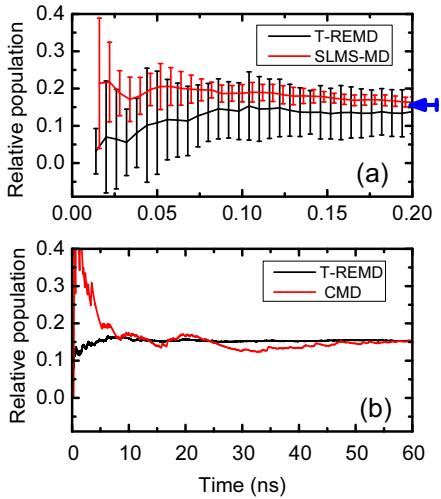}
\caption{(a) Averaged relative populations between the major basin and the
second major basin calculated by the SLMS-MD and T-REMD. The error bars are
represented by standard deviation. (b) Relative populations of the long time
single trajectory of the T-REMD and CMD simulations.}
\end{figure}

In SLMS-MD, as well as most other multiscale protocols, structure mapping
from CG to AA structures can be a major bottleneck, which is not trivial
especially for the structures far from the native state as encountered in
protein folding. Fortunately, a recent study reports significant advances in
this direction\cite{HE07}, which ensures the performance of the SLMS-MD.
Still, structure mapping may have problems in reproducing free energy
profiles because of large side-chain entropy contributions. In unfolded
structures, the same backbone conformation can accommodate many different
side-chain rotamers, which thus contributes to the side-chain entropy. In
such a case, quantitative estimate of free energy would require sampling of
multiple side-chain conformations for a single backbone conformation. It is
also worth noting that in the self-learning stage, the re-weighting formula (Eq. 3)
ensures that the final CG potential will be converged to the overall behavior
of the AA potential, which is crucial for achieving effective ResEx simulations.
However, as in other force/energy based learning methods, the contribution of the
side chain entropy will be lost to some extent, and therefore the finally derived
effective CG potential may not be identical to the real PMF.

In summary, this work presented a new idea for deriving the CG potential, and realized
accurate and efficient simulations in both the CG and AA levels. The essential idea of
this SLMS-MD is quite generic and does not depend on the specific implementation strategies
of the learning and ResEx processes. Other different ResEx strategies and learning methods,
e.g., the force match protocol, should also be applicable. Undoubtedly, the present SLMS-MD will be
benefit from any further progresses of the learning methods and the ResEx strategy, as well as
the structure mapping algorithm.

\begin{acknowledgments}
This work was partly supported by Grant-in-Aid for Scientific Research, and
partly by Research and Development of the Next-Generation Integrated
Simulation of Living Matter, a part of the Development and Use of the
Next-Generation Supercomputer Project of the Ministry of Education, Culture,
Sports, Science and Technology.
\end{acknowledgments}


\end{document}